\documentclass[aps,prl,twocolumn]{revtex4}
\usepackage{bm}
\usepackage{epsfig} 
\usepackage{amsfonts}
\usepackage{amsmath}

\allowdisplaybreaks[1] 

\newcommand*{\be}{\begin{equation}}
\newcommand*{\ee}{\end{equation}}
\newcommand*{\bea}{\begin{eqnarray}}
\newcommand*{\eea}{\end{eqnarray}}

\newcommand*{\sd}{^{\dagger}}

\def\bra#1{\langle #1|}
\def\ket#1{| #1\rangle}

\begin{document}


\title{Momentum space anisotropy in doped Mott insulators}

\author{Tiago C. Ribeiro}
\affiliation{Department of Physics, Massachusetts Institute of Technology, Cambridge, Massachusetts 02139, USA}

\date{\today}

\begin{abstract}
We study the single hole $tt't''J$-model numerically to address the 
momentum space anisotropy found in doped Mott insulators. 
A simple two band picture to understand how the doped hole is screened by the
spin background in states of different momenta is proposed.
In this picture, the disparity between the nodal and antinodal regions, 
observed by experiments in the underdoped cuprate superconductors, 
follows from local energetic considerations and amounts to the distinction 
between well defined quasiparticle behavior and spin-charge separation 
related phenomenology.
\end{abstract}

\pacs{74.72.-h}

\maketitle

\textbf{Introduction -} 
It is known, both experimentally and theoretically, that doped Mott 
insulators can be significantly anisotropic in momentum space.
Explaining this behavior is relevant to understanding the properties of 
metals close to the Mott insulating state, like the 
pseudogap regime in high-T$_c$ superconductors. \cite{PB0402,ZY0401} 

The pseudogap measured by ARPES is
the energy difference between the one-electron spectral features at the 
nodal [$\vec{k}=(\pm \tfrac{\pi}{2},\pm \tfrac{\pi}{2})$]
and antinodal points [$\vec{k}=(\pi,0),\, (0,\pi)$]. \cite{DS0373}
This difference, which increases as both the temperature and the level of 
doping in the CuO$_2$ layers are reduced, is accompanied by the anisotropic 
behavior in $k$-space present in various experimental observations.
\cite{ZY0401,DS0373,YZ0301,RS0301,KF0517,ML0491,M9965,OA0312}
Indeed, the deeply hole underdoped regime shows a clear 
nodal-antinodal dichotomy as low energy quasiparticle peaks exist around 
the nodes while there is no evidence of quasiparticle-like behavior near 
the antinodal points. \cite{ZY0401,YZ0301,RS0301}
In electron underdoped cuprates, low energy spectral weight appears
around $(\pi,0)$ instead as the 
pseudogap is pushed toward the zone diagonal. \cite{AR0201}
Interestingly, Raman spectroscopy \cite{KB0322} and the violation of the 
Wiedemann-Franz law \cite{HP0111} in electron doped materials suggests the
presence of chargeless excitations around $(\tfrac{\pi}{2},\tfrac{\pi}{2})$.
Hence, the pseudogap in the one-particle dispersion appears to
strongly reduce the electronic character of excitations 
in the pseudogap region.

Strongly correlated materials close to the Mott insulator transition
are notorious for the near degeneracy of different ordered states, 
some of which have been argued to provide a scattering mechanism
that reduces quasiparticle features in the pseudogap region.
\cite{ZY0401,ML0491,HL0401,IM9831}
In this letter, we propose a different scenario to understand the observed
momentum space anisotropy.
In particular, we argue that states close to 
$(\tfrac{\pi}{2},\tfrac{\pi}{2})$ and 
$(\pi,0)$ are different because, 
due to the competition between the spin exchange energy and 
the hole kinetic energy,
spins surrounding the hole in states near $(\tfrac{\pi}{2},\tfrac{\pi}{2})$
behave differently from the spins surrounding the hole in antinodal
states. \cite{TS0009,LL0301}
The nodal-antinodal dichotomy then reflects the existence of two distinct
ways in which the lattice spins screen the doped holes.

The 
2D $tt't''J$-model, 
which is the simplest model to study doped Mott insulators as of relevance 
to the cuprate systems, reproduces the experimental pseudogap dispersion
and accounts for anisotropic behavior in $k$-space.
\cite{TM0017,ME9916,T0417,TS0009,LL0301}
Below, we employ the exact diagonalization (ED) and the 
self-consistent Born approximation (SCBA) techniques to study the single 
hole $tt't''J$-model and to address how the hole is screened by the 
local spins in different regions of momentum space.
We find that the single hole states can be understood as the superposition 
of two distinct states, namely a hole-like quasiparticle state and a state 
where the hole strongly distorts the surrounding spins.
Due to their different properties, these states predominate in different 
parts of $k$-space in the pseudogap regime, thus leading to the disparity 
between the nodal and antinodal regions.
Our results are valid for both hole and electron doped materials --
for our purposes, the main distinction between the two regimes is 
that in the former the pseudogap opens in the antinodal region while in the 
latter it opens in the nodal region.

\textbf{Model -}
The single hole 2D $tt't''J$ Hamiltonian is
\be
H_{tt't''J} = - \sum_{\langle ij \rangle, \sigma} t_{ij} \left(
\widetilde{c}_{i,\sigma}\sd \widetilde{c}_{j,\sigma} + H.c.\right) +
\sum_{\langle ij \rangle} J_{ij} \bm{S}_i.\bm{S}_j
\label{eq:Htj}
\ee
where $t_{ij}$ equals $t$, $t'$ and $t''$ for first, second
and third nearest neighbor sites respectively and vanishes otherwise.
$\widetilde{c}_{i,\sigma}$ is the constrained operator 
$\widetilde{c}_{i,\sigma} = c_{i,\sigma} (1-n_{i,-\sigma})$. 
The exchange interaction only
involves nearest neighbor spins for which $J_{ij}=J$.
We only consider $0.2<J<0.8$ (units are set so that $t=1$).
Unless otherwise stated, all our results come from the ED of $H_{tt't''J}$ 
on a $4 \times 4$ lattice.
Since we want to analyze how the hole affects the local configuration of
the surrounding spins we believe that the study of such a small lattice
is relevant.
We also present results from the SCBA approach to the 
$tJ$ model \cite{MH9117,RV8893,RH9808}
on a $16 \times 16$ lattice to further support the ED analysis.

\begin{table}
\begin{ruledtabular}
\begin{tabular*}{\hsize}{cc|cccccc}
& J & 0.3 & 0.4 & 0.5 & 0.6 & 0.7 & 0.8 \\
\hline
& $(\tfrac{\pi}{2},\tfrac{\pi}{2})$ & 0.9994 & 0.9994 & 0.9998 & 1 & 0.9998 & 0.9990 \\ 
$t'=0$ & $(\pi,0)$ & 0.9994 & 0.9994 & 0.9998 & 1 & 0.9998 & 0.9990 \\
$t''=0$ & $(\pi,\tfrac{\pi}{2})$ & 0.9972 & 0.9977 & 0.9993 & 1 & 0.9992 & 0.9970 \\
ED & $(\tfrac{\pi}{2},0)$ & 0.9975 & 0.9980 & 0.9994 & 1 & 0.9994 & 0.9977 \\
& $(0,0)$ & 0.9946 & 0.9923 & 0.9963 & 1 & 0.9938 & 0.9766 \\ 
\hline
& $(\tfrac{\pi}{2},\tfrac{\pi}{2})$ & 0.9996 & 0.9996 & 0.9998 & 1 & 0.9998 & 0.9990 \\ 
$t'=0$ & $(\pi,0)$ & 0.9994 & 0.9994 & 0.9998 & 1 & 0.9997 & 0.9986 \\ 
$t''=0$ & $(\pi,\tfrac{\pi}{2})$ & 0.9989 & 0.9988 & 0.9996 & 1 & 0.9995 & 0.9978 \\
SCBA & $(\tfrac{\pi}{2},0)$ & 0.9989 & 0.9988 & 0.9996 & 1 & 0.9995 & 0.9978 \\
& $(0,0)$ & 0.9016 & 0.9005 & 0.9766 & 1 & 0.9842 & 0.9488 \\ 
\hline
 & $(\tfrac{\pi}{2},\tfrac{\pi}{2})$ & 0.9994 & 0.9994 & 0.9998 & 1 & 0.9998 & 0.9990 \\ 
$t'=-0.2$ & $(\pi,0)$ & -- & 1  & 0.9998 & 0.9997 & 0.9999 & 1 \\
$t''=0.1$ & $(\pi,\tfrac{\pi}{2})$ & 0.9936 & 0.9952 & 0.9986 & 1 & 0.9987 & 0.9950 \\
ED & $(\tfrac{\pi}{2},0)$ & 0.9907 & 0.9943 & 0.9986 & 1 & 0.9988 & 0.9957 \\
& $(0,0)$ & 0.9880 & 0.9856 & 0.9943 & 1 & 0.9940 & 0.9807 \\ 
\end{tabular*}
\end{ruledtabular}
\caption{\label{tab:check2d} Square of the overlap of 
$\ket{\psi_{\bm{k}},J,t',t''}$ with the Hilbert space 
$\{ \ket{\psi_{\bm{k}},J=0.2,t',t''}, \ket{\psi_{\bm{k}},J=0.6,t',t''} \}$
for different $J$ and $\bm{k}$. 
Both ED and SCBA results are shown for $t',t''=0$.
ED results are also shown for $t'=-0.2$, $t''=0.1$.
For $t'=-0.2$, $t''=0.1$ and $\bm{k}=(\pi,0)$ the Hilbert space 
$\{\ket{\psi_{\bm{k}},J=0.4,t',t''}, \ket{\psi_{\bm{k}},J=0.8,t',t''} \}$
was used instead.}
\end{table}

We start by considering the $t',t''=0$ case.
In particular, we use the ED technique to determine the lowest energy state
for each momentum, denoted by $\ket{\psi_{\bm{k}},J}$, for both 
$J=0.2$ and $J=0.6$.
Quite surprisingly, we find that if we perform the same calculation for 
different values of $J$ the resulting states $\ket{\psi_{\bm{k}},J}$
have almost complete overlap with the Hilbert space 
$\{ \ket{\psi_{\bm{k}},J=0.2}, \ket{\psi_{\bm{k}},J=0.6} \}$
(Table \ref{tab:check2d}). 
The SCBA technique leads to the same conclusion
(Table \ref{tab:check2d}), thus supporting that this result is not specific
to the $4\times 4$ lattice used in the ED calculation.

Therefore, for $0.2 < J < 0.8$, 
the wave function $\ket{\psi_{\bm{k}},J}$ can be
written approximately as 
$\ket{\psi_{\bm{k}},J} \cong q(J) \ket{Q_{\bm{k}}} + u(J) \ket{U_{\bm{k}}}$.
Here, $q(J)$ and $u(J)$ are $J$-dependent coefficients obeying the
normalization condition $q(J)^2+u(J)^2=1$.
$\ket{Q_{\bm{k}}}$ and $\ket{U_{\bm{k}}}$ are orthonormal states that 
span the 2D Hilbert space 
$\{ \ket{\psi_{\bm{k}},J=0.2}, \ket{\psi_{\bm{k}},J=0.6} \}$.
Of course, there is freedom in choosing $\ket{Q_{\bm{k}}}$ and 
$\ket{U_{\bm{k}}}$.
A physically sensible choice, though, comes from requiring that $q(J)$
increases with $J$ while $u(J)$ decreases.
Since increasing $J$ enhances the quasiparticle features of doped carriers, 
\cite{MH9117} this is automatically satisfied if the quasiparticle spectral 
weight vanishes for $\ket{U_{\bm{k}}}$, \textit{i.e.} 
$\bra{U_{\bm{k}}} \widetilde{c}_{\bm{k},\sigma} \ket{\text{HF GS}} = 0$ 
where $\ket{\text{HF GS}}$ is the groundstate of the half filled system.
This prescription uniquely determines Q states 
($\ket{Q_{\bm{k}}}$) and U states ($\ket{U_{\bm{k}}}$), \cite{OTHERJ} 
which are not eigenstates of $H_{tt't''J}$, as long as 
$\bra{\psi_{\bm{k}},J} \widetilde{c}_{\bm{k},\sigma} \ket{\text{HF GS}}\neq0$.

For general $\bm{k}$ and $t',t''$ the state 
$\ket{\psi_{\bm{k}},J,t',t''}$ only has non-zero quasiparticle spectral 
weight for $J$ larger than a certain $J_c(\bm{k},t',t'')$.
The above prescription can be applied for $J > J_c(\bm{k},t',t'')$,
in which case it is a good approximation to write
$\ket{\psi_{\bm{k}},J,t',t''} \cong q(J,t',t'') \ket{Q_{\bm{k}},t',t''} + 
u(J,t',t'') \ket{U_{\bm{k}},t',t''}$.
This fact is illustrated in Table \ref{tab:check2d} for 
$t'=-0.2$, $t''=0.1$.
Note that $0.3 < J_c < 0.4$ for $\bm{k}=(\pi,0)$, in which case we used 
the eigenstates for $J=0.4$ and $J=0.8$ to determine
$\ket{Q_{\bm{k}=(\pi,0)},t',t''}$ and $\ket{U_{\bm{k}=(\pi,0)},t',t''}$. 
As discussed below, for $J<J_c(\bm{k},t',t'')$ the eigenstate 
$\ket{\psi_{\bm{k}},J,t',t''}$ displays properties akin to those of U states
and we denote it by $\ket{\widetilde{U}_{\bm{k}},J,t',t''}$ 
(where the tilde is used to distinguish it from the U state
derived for $J>J_c(\bm{k},t',t'')$, which has no $J$ dependence).

The previous argument points out that the eigenstates 
$\ket{\psi_{\bm{k}},J,t't,''}$ 
have features that get enhanced by an increasing $J$ and features
that become more pronounced when $J$ is reduced.
By definition, Q and U states capture these features and,
not surprisingly, below we show they display distinct physical properties:
in Q states the spin antiferromagnetic (AF) order is robust to the presence 
of the hole while in U states the spins rearrange to facilitate hole hopping.
On general grounds, a construction similar to the 
above one might be valid for models other than the 2D $tt't''J$-model.
The significant fact about this model is that, for experimentally 
relevant parameters, the overlap of both Q and U states with 
$\ket{\psi_{\bm{k}},J,t',t''}$ is large and exhibits a 
considerable momentum dependence which leads to the anisotropy
in $k$-space. \cite{3D}

\begin{figure}
\includegraphics[width=0.48\textwidth]{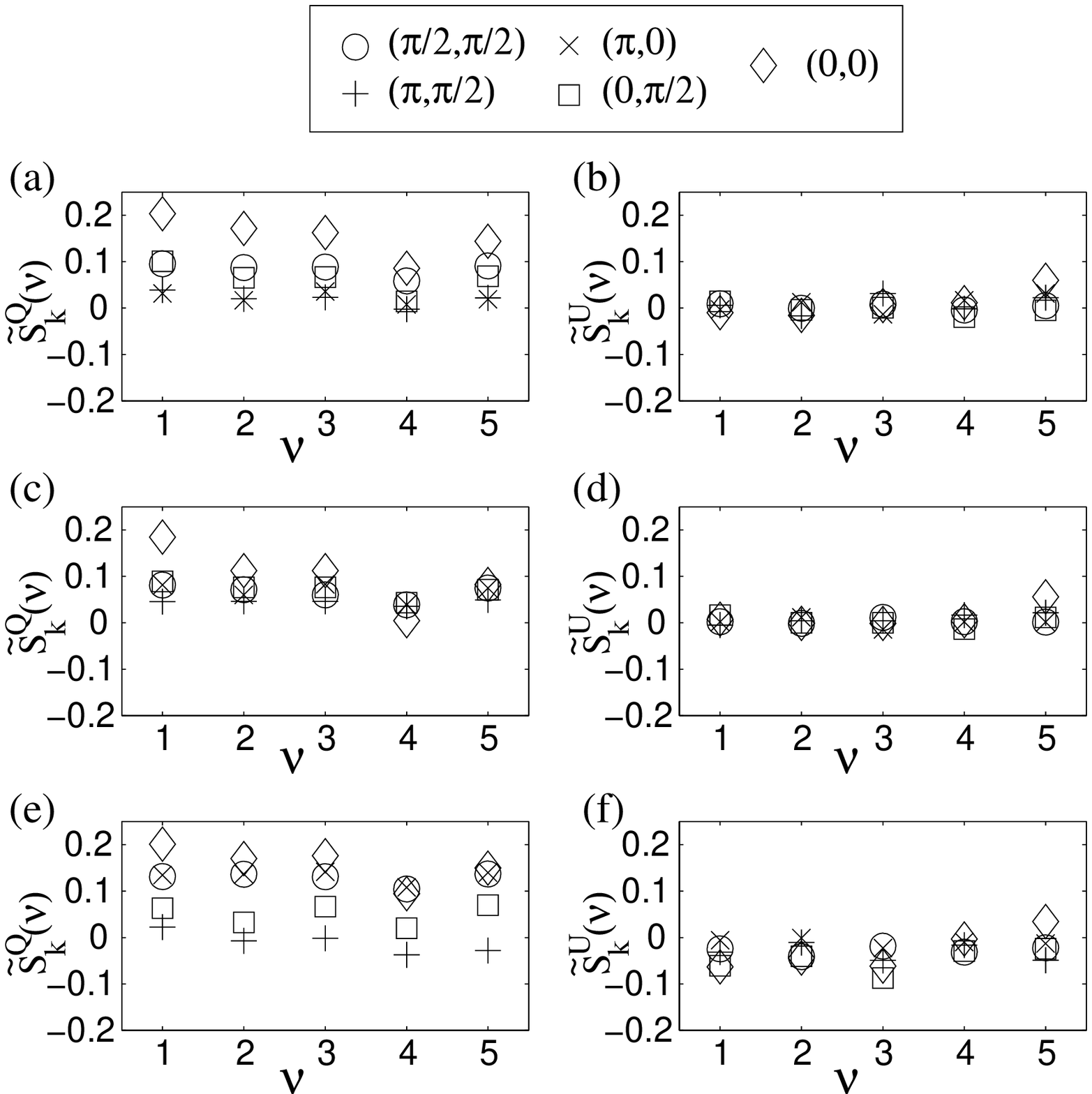}
\caption{\label{fig:RK}
Average value of the staggered magnetization on sites that are $\nu^{th}$ 
nearest neighbors to the hole, 
$\widetilde{S}_{\bm{k}}^Q(\nu)\equiv
\langle(-)^{i_x+i_y}S_{\bm{k}}^Q(\bm{i})\rangle_{\nu}$ 
and $\widetilde{S}_{\bm{k}}^U(\nu)\equiv
\langle(-)^{i_x+i_y}S_{\bm{k}}^U(\bm{i})\rangle_{\nu}$, 
for different momenta $\bm{k}$.
(a) and (b) $t'=-0.3, t''=0.2$. 
(c) and (d) $t',t''=0$.
(e) and (f) $t'=0.3, t''=-0.2$. 
}
\end{figure}

\textbf{Properties of Q and U states - }
To explore 
how the hole affects the spin background in Q and U states consider
the spin density pattern around the hole, 
$S_{\bm{k}}^Y(\bm{i})\equiv\bra{Y_{\bm{k}}} \sum_j S_{j+i}^z 
\widetilde{c}_{j,+1/2} \widetilde{c}_{j,+1/2}\sd \ket{Y_{\bm{k}}} + 
\tfrac{1}{N-1}\tfrac{1}{2}$ with $Y=Q,U$, \cite{MAGNETIZATION}
as well as the hole momentum distribution function, 
$n_{\bm{k}}^Y(\bm{q},\sigma)\equiv\bra{Y_{\bm{k}}} 
\widetilde{c}_{-\bm{q},-\sigma} \widetilde{c}_{-\bm{q},-\sigma}\sd  
\ket{Y_{\bm{k}}}$ for $Y=Q,U$.
Figs. \ref{fig:RK}(a), \ref{fig:RK}(c) and \ref{fig:RK}(e) depict
the staggered magnetization in Q states, $(-)^{i_x+i_y}S_{\bm{k}}^Q(\bm{i})$,
for different $\bm{k}$, $t'$ and $t''$.
Since these states have a well defined quasiparticle character, the
doped hole coexists with the AF 
spin pattern inherited from the undoped system.
For the same reason, the hole momentum distribution function
$n_{\bm{k}}^Q(\bm{q},-\tfrac{1}{2})$ is peaked at 
$\bm{q}=\bm{k}$ while a smaller peak is also observed at 
$\bm{q}=\bm{k}+(\pi,\pi)$ due to the strong AF correlations. \cite{EO9541}

\begin{figure}
\includegraphics[width=0.48\textwidth]{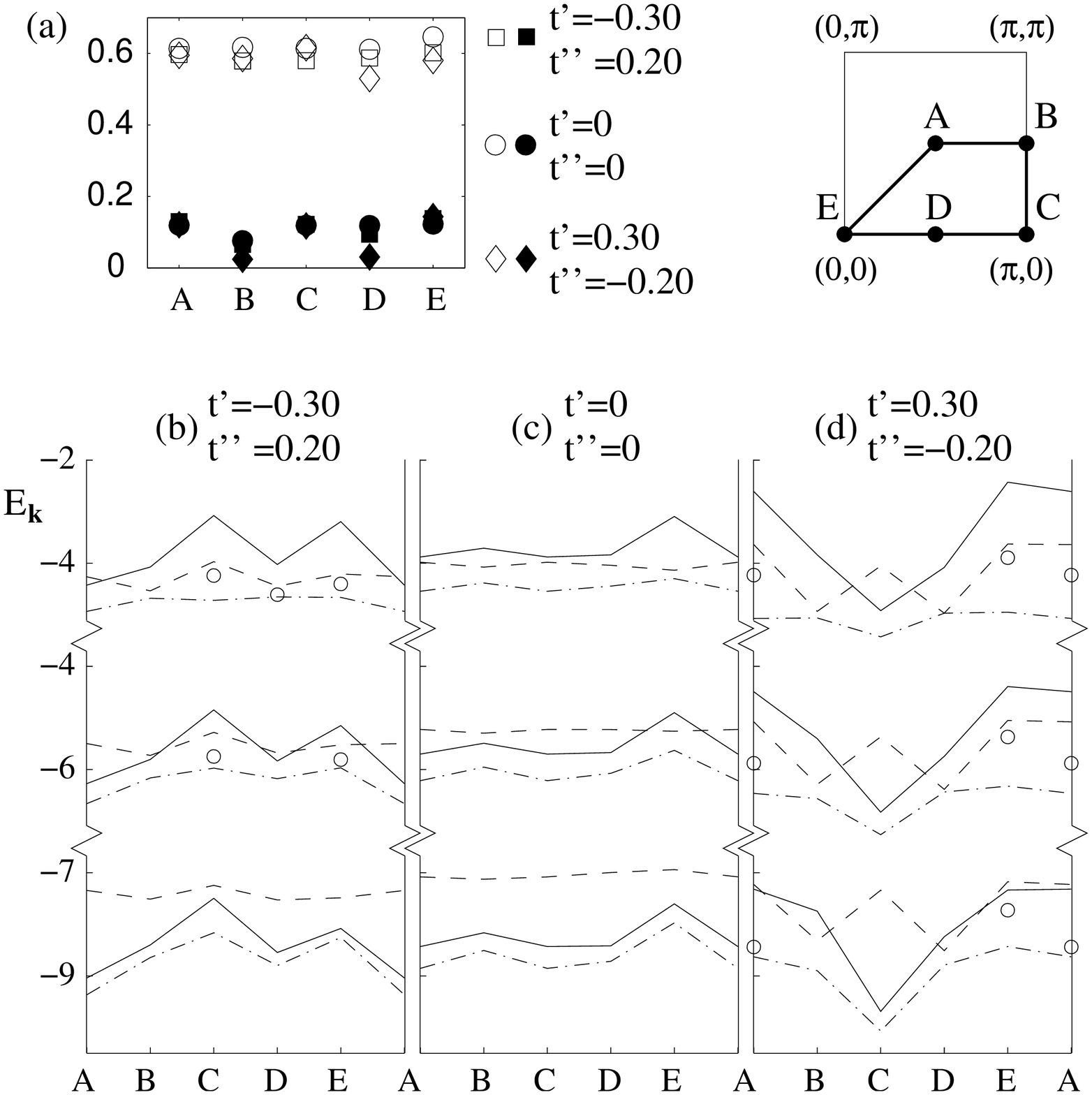}
\caption{\label{fig:bands}
(a) 
$\sum_{\bm{q}}' n_{\bm{k}}^U(\bm{q},-\tfrac{1}{2})$.
Empty symbols involve sum over $\bm{q}=(\pi,\pi)$,
$\bm{q}=(\pm\tfrac{\pi}{2},\pi)$ and $\bm{q}=(\pi,\pm\tfrac{\pi}{2})$.
Full symbols involve sum over $\bm{q}=(0,0)$,
$\bm{q}=(\pm\tfrac{\pi}{2},0)$ and $\bm{q}=(0,\pm\tfrac{\pi}{2})$.
(b)-(d) Dispersion relations for $\ket{Q_{\bm{k}}}$
(full line), $\ket{U_{\bm{k}}}$ (dashed line) and $\ket{\psi_{\bm{k}}}$
(dash-dot line). Upper, middle and lower set of dispersions are obtained
for $J$ equal to $0.2$, $0.4$ and $0.7$ respectively.
($\circ$) indicates the best energy obtained by a linear combination of
$\ket{Q_{\bm{k}}}$ and $\ket{U_{\bm{k}}}$ when $J<J_c(\bm{k},t',t'')$ 
(in which case $\ket{\psi_{\bm{k}}} = \ket{\widetilde{U}_{\bm{k}},J,t',t''}$).
}
\end{figure}

U states have no quasiparticle weight and display quite different behavior.
Indeed, Figs. \ref{fig:RK}(b), \ref{fig:RK}(d) and \ref{fig:RK}(f)
show that in $\ket{U_{\bm{k}}}$ the AF spin pattern of the
undoped system is destroyed and the staggered magnetization 
around the hole [given by $(-)^{i_x+i_y}S_{\bm{k}}^U(\bm{i})$] 
is very close to zero and even negative.
Moreover, the hole momentum distribution function 
$n_{\bm{k}}^U(\bm{q},-\tfrac{1}{2})$ peaks around 
$\bm{q}=(\pi,\pi)$ for all momenta $\bm{k}$ [Fig. \ref{fig:bands} (a)].
The same real and momentum space properties 
were checked to hold for $\widetilde{\text{U}}$ states [these are
the energy eigenfunctions $\ket{\psi_{\bm{k}},J,t',t''}$ when
$J<J_c({\bm{k}},t',t'')$].
This fact is illustrated for $\widetilde{\text{U}}$ states concerning
the parameter regime relevant to both hole doped cuprates
($J=0.4$, $t'=-0.3$, $t''=0.2$) and electron doped cuprates
($J=0.4$, $t'=0.3$, $t''=-0.2$) \cite{TM0017} in Table \ref{tab:MD},
which explicitly shows that in these states the hole density also peaks 
around $(\pi,\pi)$ independently of the momenta $\bm{k}$.

According to the above spin density results the extra 
$S^z=-\tfrac{1}{2}$ spin introduced by doping spreads away from the vacancy
in both U and $\widetilde{\text{U}}$ states.
The resulting loss of spin exchange energy is accompanied by a gain in 
hole kinetic energy.
Indeed, the hole momentum distribution results support that, in these
states, the hole always lies around the bare band bottom 
[which is located at $(\pi,\pi)$]. 
This evidence resembles predictions from spin-charge separation scenarios. 
Indeed, within the slave-boson formalism,
the electron decays into a charged spinless holon, which condenses at 
$(\pi,\pi)$, and a chargeless spinon, which describes the delocalized 
spin-$\tfrac{1}{2}$ that carries the remaining momentum.
We should remark that our calculation involves equal time properties
in a small lattice and does not aim to prove the existence of true 
spin-charge separation.
However, it supports that in U and $\widetilde{\text{U}}$ states the 
lattice spins screen the hole in conformity with 
short range aspects of spin-charge separation phenomenology.

\begin{table}
\begin{ruledtabular}
\begin{tabular*}{\hsize}{c|cc|cc|}
& \multicolumn{2}{c|}{$\!\!\!\!\! t'=-0.3; t''=0.2$} &
\multicolumn{2}{c|}{$\!\!\!\!\! t'=0.3; t''=-0.2$} \\
$\bm{k}$ & $(\pi,0)$ & $(0,0)$ & $(\tfrac{\pi}{2},\tfrac{\pi}{2})$ & $(0,0)$ \\
$\bm{q}=(0,0)$ & 0.0329 & 0.0343 & 0.0164 & 0.0328 \\
$\bm{q}=(\pi,\pi)$ & 0.5437 & 0.6051 & 0.5293 & 0.5141 \\
\end{tabular*}
\end{ruledtabular}
\caption{\label{tab:MD} 
$\sum_{\bm{q}}' n_{\bm{k}}^{\widetilde{U}}(\bm{q},-\tfrac{1}{2})\equiv
\sum_{\bm{q}}'\bra{\widetilde{U}_{\bm{k}}}\widetilde{c}_{-\bm{q},\tfrac{1}{2}} 
\widetilde{c}_{-\bm{q},-\tfrac{1}{2}}\sd  \ket{\widetilde{U}_{\bm{k}}}$.
$\bm{q}=(0,0)$ results involve sum over $\bm{q}=(0,0)$,
$\bm{q}=(\pm\tfrac{\pi}{2},0)$ and $\bm{q}=(0,\pm\tfrac{\pi}{2})$.
$\bm{q}=(\pi,\pi)$ results involve sum over $\bm{q}=(\pi,\pi)$,
$\bm{q}=(\pm\tfrac{\pi}{2},\pi)$ and $\bm{q}=(\pi,\pm\tfrac{\pi}{2})$.
$J=0.4$.}
\end{table}

\begin{table}
\begin{ruledtabular}
\begin{tabular*}{\hsize}{ccccccc}
$t'$ & $t''$ & $\Delta E^{\psi}$ & $\Delta E^Q$ & $\Delta E^U$ &
$W_{\bm{k'}}^Q$ & $W_{\bm{k''}}^Q$ 
\\
\hline
-0.3 & 0.2 & 0.69 & 1.43 & 0.22 & 0 & 0.75 \\
-0.2 & 0.1 & 0.56 & 0.92 & 0.14 & 0.45 & 0.72 \\
0 & 0 & 0 & 0 & 0 & 0.66 & 0.66 \\
0.2 & -0.1 & -0.75 & -1.08 & -0.10 & 0.76 & 0.50 \\
0.3 & -0.2 & -0.80 & -2.33 & -0.29 & 0.82 & 0 \\
\end{tabular*}
\end{ruledtabular}
\caption{\label{tab:pseudogap}
$\Delta E^Q$, $\Delta E^U$, $\Delta E^{\psi}$ and
$W_{\bm{k}}^Q$ with $\bm{k}=\bm{k'} \equiv (\pi,0)$
and $\bm{k}=\bm{k''} \equiv (\tfrac{\pi}{2},\tfrac{\pi}{2})$ for several 
$t'$ and $t''$ and $J=0.4$.}
\end{table}

The previous results demonstrate that the local configuration of
spins encircling the hole is quite different in Q and U states.
In this context, it is important to remark that the influence of $t',t''$ 
on the hole dispersion is sensitive to the surrounding spin environment.
For instance, intrasublattice hopping is not frustrated by AF correlations
and, indeed, the hole in Q states (which is surrounded by a spin configuration
reminiscent of the undoped AF groundstate) strongly disperses 
along the 
AF Brillouin zone boundary (AFBZB) 
for $t',t''\neq0$ (see Fig. \ref{fig:bands}).
The opposite limit occurs, for example, in certain $U(1)$ spin liquids, 
where the underlying spin correlations inhibit coherent intrasublattice 
hopping. \cite{RW0301}
This limit is closer to what is observed in U states, as
$t',t''$ are heavily renormalized by the spin background (whose
AF correlations are depleted by the hole nearest-neighbor hopping
processes).
In fact, Table \ref{tab:pseudogap} explicitly shows that 
$\Delta E^Q\equiv E_{(\pi,0)}^Q-E_{(\pi/2,\pi/2)}^Q$ is almost one order
of magnitude larger than $\Delta E^U\equiv E_{(\pi,0)}^U-E_{(\pi/2,\pi/2)}^U$,
where
$E_{\bm{k}}^Q\equiv\bra{Q_{\bm{k}}}H_{tt't''J}\ket{Q_{\bm{k}}}$ and
$E_{\bm{k}}^U\equiv\bra{U_{\bm{k}}}H_{tt't''J}\ket{U_{\bm{k}}}$.

\textbf{Momentum space anisotropy - }
In the $tt't''J$-model the intrasublattice hopping parameters $t',t''$
control the dispersion 
$E_{\bm{k}}^{\psi}\equiv\bra{\psi_{\bm{k}}}H_{tt't''J}\ket{\psi_{\bm{k}}}$
along the AFBZB 
and, thus, the pseudogap energy as well (this is the energy difference 
between the antinodal and nodal states 
$\Delta E^{\psi}\equiv E_{(\pi,0)}^{\psi}-E_{(\pi/2,\pi/2)}^{\psi}$).
Indeed, both Fig. \ref{fig:bands} and Table \ref{tab:pseudogap} show that
these parameters set the magnitude of the pseudogap,
as well as its location in $k$-space. \cite{T0417,TY0403}

Interestingly $t',t''$ also control the difference in the quasiparticle 
character of $(\pi,0)$ and $(\tfrac{\pi}{2},\tfrac{\pi}{2})$ states, 
as supported by the dependence of the overlap integral
$W_{\bm{k}}^Q\equiv |\langle \psi_{\bm{k}}\ket{Q_{\bm{k}}}|^2$
on these parameters (Table \ref{tab:pseudogap}).
The resulting $t',t''$ driven momentum space anisotropy is 
particularly evident when both Q and U states have large overlaps with 
the eigenstates $\ket{\psi_{\bm{k}}}$, as it is the case for 
the value of $J$ which is relevant to the cuprate systems, namely $J=0.4$.
\cite{TM0017,JDEPENDENCE}

The main message of this letter is that the above roles of $t',t''$
can be understood with a simple two band picture involving
Q and U states which captures the screening properties of the local 
spins encircling the hole.
Indeed, as previously discussed, 
intrasublattice hopping is frustrated in U states 
and the energies $E_{(\pi,0)}^U$ and $E_{(\pi/2,\pi/2)}^U$
have a small dependence on $t',t''$.
In contrast, $t'\!<\!0$ and $t''\!>\!0$ clearly increase the energy 
$E_{(\pi,0)}^Q$ and clearly decrease $E_{(\pi/2,\pi/2)}^Q$ 
(Fig. \ref{fig:bands}).
Since $\ket{\psi_{\bm{k}}}$ is a superposition of $\ket{Q_{\bm{k}}}$
and $\ket{U_{\bm{k}}}$, these $t',t''$ induced changes in
the energy of Q states
push $E_{(\pi,0)}^{\psi}$ upward and lower $E_{(\pi/2,\pi/2)}^{\psi}$.
Therefore, they increase the pseudogap energy $\Delta E^{\psi}$.
These changes also increase the energy difference 
$E_{(\pi,0)}^Q - E_{(\pi,0)}^U$ while reducing
$E_{(\pi/2,\pi/2)}^Q - E_{(\pi/2,\pi/2)}^U$ and, as a result,
$W_{(\pi,0)}^Q$ and $W_{(\pi/2,\pi/2)}^Q$ become smaller and
larger respectively.
For $J=0.4, t'=-0.3, t''= 0.2$ 
the energy $E_{(\pi,0)}^Q$ is so large that the minimum energy obtained 
by a linear combination of $\ket{Q_{(\pi,0)}}$ and $\ket{U_{(\pi,0)}}$ 
becomes higher than that of a different state $\ket{\widetilde{U}_{(\pi,0)}}$. 
As a result, $W_{(\pi,0)}^Q = 0$.
Since $W_{(\pi/2,\pi/2)}^Q = 0.75$, the $(\tfrac{\pi}{2},\tfrac{\pi}{2})$
state has strong quasiparticle character and a sharp disparity between
nodal and antinodal states is encountered for these parameter values.
Similar behavior is obtained when $t'\!>\!0$ and $t''\!<\!0$ 
with the role of momenta $(\pi,0)$ and $(\tfrac{\pi}{2},\tfrac{\pi}{2})$ 
interchanged (Table \ref{tab:pseudogap}).

The above argument shows that, in the pseudogap region, the kinetic energy 
of a hole surrounded by an AF configuration of spins increases as the 
pseudogap gets larger.
When this kinetic energy exceeds a certain value the spins around the 
hole lose their staggered pattern and, at short distances, the doped
spin and charge separate.
The resulting nodal-antinodal dichotomy,
found to occur in the parameter regimes concerning 
hole and electron doped cuprates, 
follows from local energetic considerations and not from the interaction
of holes with other holes, an incipient order parameter or disorder.
It also agrees with previous numerical evidence for spin-charge 
separation phenomenology at high energy. \cite{TS0009,LL0301}

Following our results, a new mean field theory of the $tt't''J$-model was 
recently proposed that describes holes dressed by two different spin
configurations. \cite{RW0450}
It accounts for the observed momentum space anisotropy and is in 
good agreement with the doping dependence of ARPES experiments.

\begin{acknowledgments}
The author is grateful to X.-G. Wen and P.A. Lee for discussions 
and valuable comments on the manuscript. 
This work was partially supported by the Funda\c c\~ao 
Calouste Gulbenkian Grant No. 58119 (Portugal), 
NFSC Grant No. 10228408 (China), NSF Grant No. DMR-01-23156 and
NSF-MRSEC Grant No. DMR-02-13282.

\end{acknowledgments}

\newcommand*{\PR}[1]{Phys.\ Rev.\ {\textbf {#1}}}
\newcommand*{\PRL}[1]{Phys.\ Rev.\ Lett.\ {\textbf {#1}}}
\newcommand*{\PRB}[1]{Phys.\ Rev.\ B {\textbf {#1}}}
\newcommand*{\PTP}[1]{Prog.\ Theor.\ Phys.\ {\textbf {#1}}}
\newcommand*{\MPL}[1]{Mod.\ Phys.\ Lett.\ {\textbf {#1}}}
\newcommand*{\JPC}[1]{Jour.\ Phys.\ C {\textbf {#1}}}
\newcommand*{\RMP}[1]{Rev.\ Mod.\ Phys.\ {\textbf {#1}}}
\newcommand*{\RPP}[1]{Rep.\ Prog.\ Phys.\ {\textbf {#1}}}
\newcommand*{\PHY}[1]{Physics {\textbf {#1}}}
\newcommand*{\ZP}[1]{Z.\ Phys.\ {\textbf {#1}}} 
\newcommand*{\JETP}[1]{Sov.\ Phys.\ JETP Lett.\ {\textbf {#1}}}
\newcommand*{\PLA}[1]{Phys.\ Lett.\ A {\textbf {#1}}}
\newcommand*{\AP}[1]{Adv.\ Phys.\ {\textbf {#1}}}
\newcommand*{\JLTP}[1]{J.\ Low Temp.\ Phys.\ {\textbf {#1}}}
\newcommand*{\SC}[1]{Science\ {\textbf {#1}}}
\newcommand*{\NA}[1]{Nature\ {\textbf {#1}}}
\newcommand*{\CMAT}[1]{cond-mat/{#1}}
\newcommand*{\JPSJ}[1]{J.\ Phys.\ Soc.\ Jpn.\ {\textbf {#1}}}
\newcommand*{\PC}[1]{Physica.\ C {\textbf {#1}}}
\newcommand*{\JPCS}[1]{J.\ Phys.\ Chem.\ Solids\ {\textbf {#1}}}
\newcommand*{\APNY}[1]{Ann.\ Phys.\ (N.Y.) {\textbf {#1}}}
\newcommand*{\SSC}[1]{Solid State Commun.\ {\textbf {#1}}}
\newcommand*{\SST}[1]{Supercond.\ Sci.\ Technol.\ {\textbf {#1}}}
\newcommand*{\PRPT}[1]{Phys.\ Rep.\ {\textbf {#1}}}


\begin{thebibliography}{99}
\bibitem{3D}
We expect that U states derived for the 3D $tt't''J$-model be of less
relevance than those obtained for the 
2D model since AF 
correlations are more robust in 3D.
\bibitem{MAGNETIZATION}
To reduce finite size effects we subtract the average magnetization
$\tfrac{1}{N-1}\sum_j \langle S_j^z \rangle = -\tfrac{1}{N-1}\tfrac{1}{2}$, 
where $N$ is the number of lattice sites.
\bibitem{JDEPENDENCE}
$J$ increases the robustness of the AF background thus 
enhancing the quasiparticle behavior.
Hence, larger values of $J$ decrease the weight of U states.
\end{thebibliography}
\end{document}